\documentclass[aps,prb,twocolumn,reprint,superscriptaddress]{revtex4-2}

\usepackage[caption=false]{subfig}
\newcommand{\phantomsubfloat}[1]{
    {
        \captionsetup[subfloat]{farskip=0pt,captionskip=0pt}
        \captionsetup[subfigure]{labelformat=empty}
        \subfloat{#1}
    }%
}

\usepackage{graphicx}
\usepackage{dcolumn}
\usepackage{bm}
\usepackage{float} 
\usepackage[table,xcdraw]{xcolor}
\DeclareUnicodeCharacter{2212}{-} 
\usepackage{amsmath}
\usepackage{ulem}
\usepackage{color}
\usepackage{xcolor}

\usepackage{array}
\usepackage{multirow}

\usepackage{hyperref}

\hypersetup{
    hidelinks,
    colorlinks=true,
    linkcolor=blue,
    filecolor=blue,      
    urlcolor=blue,
    citecolor=blue
}

\begin{document}

\preprint{APS/123-QED}

\title{Revisiting the Extraction of Coupling Strength for Polaron Hopping from $ab~initio$ Approach} 

\author{Hala~Houmsi}
\email{Hala.Houmsi@cea.fr}
\affiliation{Univ. Grenoble Alpes, CEA, Leti, F-38000, Grenoble, France}
\affiliation{Lynred, 364 Avenue de Valence, 38113,Veurey-Voroize, France}

\author{Benoit~Skl\'enard}
\affiliation{Univ. Grenoble Alpes, CEA, Leti, F-38000, Grenoble, France}

\author{Marc~Guillaumont}
\affiliation{Lynred, 364 Avenue de Valence, 38113,Veurey-Voroize, France}

\author{François~Triozon}
\affiliation{Univ. Grenoble Alpes, CEA, Leti, F-38000, Grenoble, France}
  
\author{Jing~Li}
\email{Jing.Li@cea.fr}
\affiliation{Univ. Grenoble Alpes, CEA, Leti, F-38000, Grenoble, France}

\date{\today}

\begin{abstract}
Accurately determining the coupling strength between polaron states is essential to describe charge-hopping transport in materials. In this work, we revisit methodologies to extract coupling strengths using $ab~initio$ approaches. Our findings underscore the critical role of incorporating anharmonic effects in the model Hamiltonian  when analyzing total energy variations along reaction coordinates. Furthermore, we demonstrate that coupling strength extraction based on total energy calculated from $ab~initio$ approaches is fundamentally limited due to the stabilization of diabatic states, which do not involve coupling strength. We demonstrate that such limitation exists in both DFT+U and HSE hybrid functionals, which are widely used in the study of polaron transport. Instead, we suggest extracting coupling strength directly from the electronic structure of a system in neutral conditions. The neutral condition avoids overestimating coupling strength due to additional energy splitting between bonding and anti-bonding states from the charging energy. This study highlights the limitations of existing methods and introduces a robust framework for accurately extracting coupling parameters, paving the way for improved modeling of charge transport in complex materials.
\end{abstract}

\maketitle


\section{Introduction}
Strong electron-phonon coupling facilitates the localization of an electron, accompanied by a local structural distortion, leading to the formation of a polaron state.
Unlike band-like transport, small polaron transport is the propagation of the localized charge accompanied by a structural distortion which can be described by a reaction coordinate.
Charge transfer theories, such as Marcus theory \cite{marcus_electron_1985} and Landau-Zener theory \cite{newton_electron_1984,Zener_1932} represent the fundamental theoretical frameworks traditionally employed to quantify the Charge Transfer (CT) rate, where the key parameters are the activation energy and the coupling strength.

\begin{figure}
\includegraphics[width=\columnwidth]{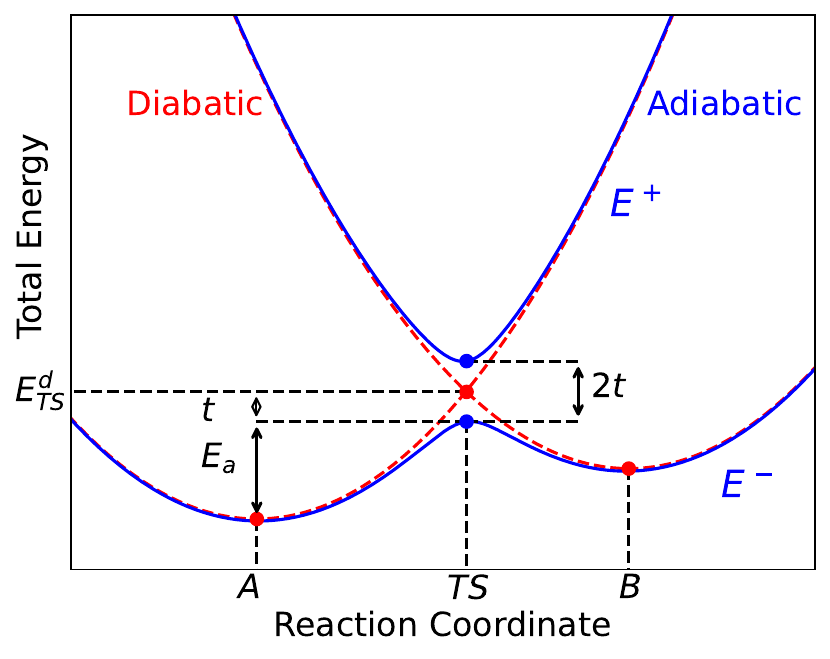}
\caption{
Typical total energy profile of polaron hopping from the initial state A to the final state B along the reaction coordinate. The activation energy $E_a$, the coupling strength $t$ between the two states, the ground and excited adiabatic energy surfaces $E^−$ and $E^+$, the transition state (TS) and the diabatic intersection $E^d_{TS}$ are indicated.
}
\label{fig:0}
\end{figure}

The typical total energy profile used to describe CT is depicted in Fig.~\ref{fig:0}. It is represented by a reaction coordinate linearly projected along a vector connecting the initial state A to the final state B. The energy surface exhibits local minima at both the initial and final states.
When moving away from an energy local minimum along the reaction coordinate, the energy is generally assumed to increase quadratically under the harmonic approximation of atomic oscillations.
When the reaction coordinate varies fast enough (relative to the motion of the electron) such that the localized electron remains at its original site, (namely the diabatic limit), the two energy surfaces cross each other ($E^d_{TS}$ see red curves in Fig.~\ref{fig:0}) \cite{small_theory_2003}.
If the change of reaction coordinate is slower than the electron motion (adiabatic limit), the system follows the ground state energy surface.
At the transition, where the two polaron states become resonant, the system adopts the bonding state configuration with lower energy, due to the coupling between those two polaron states. 
Mathematically, using the normalized reaction coordinates, $x \in [0,1]$, which spans from the initial to the final state, the two polaron states under harmonic approximation are expressed as: 
\begin{eqnarray}
    E_A &=& \lambda_A x^2 \label{eq:parabolic_Ea},\\
    E_B &=& \lambda_B (1-x)^2 + \Delta G \label{eq:parabolic_Eb},
\end{eqnarray}
where $\lambda$ is the reorganization energy of a polaron state, and $\Delta G$ the total energy difference between final and initial states \cite{marcus_electron_1985}.
Taking into account the coupling strength, the adiabatic ground state $E^-$ and excited state $E^+$ are simply the eigenvalues of the following Hamiltonian: 
\begin{equation}
\textbf{$H_{tot}$}=
\begin{bmatrix}
E_A & t \\
t & E_B 
\end{bmatrix},
\label{eq:H_model}
\end{equation}
where $t$ is the coupling strength. 

Based on the simple two-site model (Eq.~\ref{eq:H_model}), the coupling strength between polaron states can be evaluated by different approaches, mainly using the energies computed from $ab~initio$ calculations.  
For example, the ground state energy ($E^-$) is accessible from $ab~initio$ methods along the reaction coordinate. By fitting $ab~initio$ data with the lower eigenvalue of the model, key parameters such as the reorganization energy, activation energy, and coupling strength are extracted in one-shot \cite{defrance_ab_2022}.
Another approach focuses on the Transition State (TS) by obtaining the adiabatic ground and excited states. The energy difference yields twice the coupling strength, $2t=E^+_{TS}-E^-_{TS}$. This relationship can be easily demonstrated using equation \ref{eq:H_model}. At the crossing point, the diabatic states are degenerate yielding $E_A=E_B=E$. The eigenvalues of this new Hamiltonian are $E^\pm=E\pm t$. The difference between the two eigenvalues ($E^+$ and $E^-$) gives twice the coupling strength. An alternative approach is to access the two-fold degenerate diabatic energy ($E^{d}_{TS}$) and the ground adiabatic state at the transition state, namely $t=E^{d}_{TS}-E^-_{TS}$. 
The diabatic energy at the transition state can be estimated either by fitting the two parabolas close to local minima and extrapolating for the intersection at TS \cite{carey_hole_2021,chen_tuning_2023} or by using constrained DFT, which enforces the electron localization at its original site \cite{wu_direct_2006, kaduk_constrained_2012, McKenna_2015, Wu_2018}.

Besides total energy, the coupling strength can also be extracted from the electronic structure. The polaron state typically falls inside the bandgap. At TS, two polaron states form Bonding (BS) and Anti-Bonding States (ABS) in the electronic structure. The energy difference yields twice the coupling strength \cite{wang_exploring_2016,watthaisong_transport_2019, wang_exploring_2016, palermo_migration_2024, falletta_polaron_2023}. Similarly to the total energy method, the transition of the polaron state in the band gap can be described using the following Hamiltonian: 
\begin{equation}
\textbf{$H_{sp}$}=
\begin{bmatrix}
\epsilon_A & t \\
t & \epsilon_B 
\end{bmatrix},
\label{eq:Hprime_model}
\end{equation}
where $t$ is the coupling constant and $\epsilon$ is the single particle energy level of the polaron state. 
The band structure, as well as the total energy method mentioned above, should yield identical values for the coupling strength within the approximations of the two-site model as presented by Wang et al. \cite{wang_application_2024}. 

For polaron states mediated by defects, a way to extract coupling strength is from the polaron states of a single and double defects in electronic structure \cite{Hirchaou_2024}. The energies of single defect states give onsite energies and the energies of double defect states contain the coupling strength. Such a method avoids searching for the resonant condition in reaction coordinates and is efficient and suitable to study charge hopping in amorphous structures.

Quantifying the coupling strength by combining the model Hamiltonian  and $ab~initio$ methods is plausible. However, caution must be paid to the approximations made in the model. 
In addition, to stabilize the polaron state, different methods including DFT+U, Hybrid functional, and constrained DFT are commonly employed. How do these techniques influence the energy profiles? Does the physical picture depicted in Fig.~\ref{fig:0} remain valid for the coupling strength extraction? The extraction from electronic structure is typically performed in charged system. To what extent the energy splitting corresponds to the coupling remains to be discussed.
In this work, we revisit different approaches to extract the coupling strength, reveal their correctness and limitations, and suggest general guidelines for coupling strength extraction. 

The remainder of this article is organized as follows: Section \ref{sec:Method} describes the computational methods used in this study. Section \ref{sec.anharmonicity} presents the influence of anharmonicity on coupling strength extraction in total energy approaches. Sections \ref{sec.DFT+U} and \ref{sec.HSE} respectively provide DFT+U and HSE06 results of three polaron jumps in V$_2$O$_5$. In those sections, we confront total energy and band structure methods for the coupling strength extraction from $ab~initio$ data based on the two site Hamiltonian models \textbf{$H_{tot}$} and \textbf{$H_{sp}$} . Section \ref{sec.discuss} contains the discussion about our findings in the context of existing literature. Section \ref{sec:conclusion} concludes the study with key takeaways.

\section{Method}\label{sec:Method}
For demonstration purposes, we choose a widely studied material, orthorhombic V$_2$O$_5$ (spacegroup Pmmn) which is an interesting transition metal oxide for many applications, such as energy storage~\cite{hu_vanadium_2023}, photo catalyst~\cite{Jayaraj_2018}, electrochromic devices~\cite{Le_2022}, and gas sensor~\cite{Alrammouz_2021}. We employ the Projector Augmented Wave method (PAW)~\cite{blochl_projector_1994} with plane-waves basis set, as implemented in the Vienna Ab initio Simulation Package (VASP) ~\cite{Kresse_1993, Kresse_1996a, Kresse_1996b, Kresse_1999,hafner_ab-initio_2008}. DFT calculations were conducted on $1\times4\times3$ supercell ($168$ atoms) with a cut-off energy of $400$~eV, and the k-point mesh was restricted to the $\Gamma$-point due to the large cell size. In the electronic self-consistent loop, the total energy is converged within $10^{−8}$~eV. Concerning structural relaxation, the ionic positions were fully optimized, achieving forces on atoms below $0.01$~eV/\AA. 

We have performed two types of DFT calculation in this study, DFT+U and HSE06. For DFT+U, We chose the Generalized Gradient Approximation (GGA) functional developed by Perdew, Burke, and Ernzerhof (PBE) \cite{perdew_generalized_1996}. To account for the strong electron correlations in the vanadium 3d orbitals, we apply the on-site Hubbard term (U) in the DFT calculation (DFT+U) \cite{anisimov_band_1991, anisimov_first-principles_1997, baeriswyl_hubbard_1995, hubbard_electron_1963}, via the Dudarev implementation \cite{dudarev_electron-energy-loss_1998} with the effective $U$ value set to $3.5$~eV. This value was determined based on the scheme developed by Falleta et al. \cite{falletta_hubbard_2022} accompanied with FWP (Falleta, Wiktor, Pasquarello) corrections \cite{falletta_finite-size_2020} (See Appendix \ref{detU} for details). Section \ref{sec.HSE} uses HSE06 range-separated hybrid functional. The latter combines 25\% exact exchange with 75\% PBE exchange in the short range, while retaining full PBE exchange in the long range. By mitigating Self-Interaction Errors (SIE), it often yields a more accurate description of the electronic structure of materials \cite{HSE_2006}. Parameters in HSE06 functional are kept as default values. 

To stabilize an electron-polaron state, i.e., to localize an electron on a specific vanadium atom, we applied a $10\%$ distortion (expansion) to the neutral bulk structure in order to generate an initial structure. This is done by selectively altering the bond distances between a chosen vanadium atom and its nearest oxygen neighbors within a cut-off radius of $3$~\AA. Then, we added an electron and relaxed the atomic positions to obtain a polaron state. The same procedure was applied to stabilize the polarons located on neighboring atoms.
Atomic structures along the polaron hopping path were generated using the linear interpolation method between the initial and final atomic structures.
$Ab~initio$ calculation could converge to the ground or excited states depending on the initial conditions, such as initial magnetic moments and charge density. 

\section{Anharmonicity}
\label{sec.anharmonicity}

\begin{figure*}
\includegraphics[width=2.05\columnwidth]{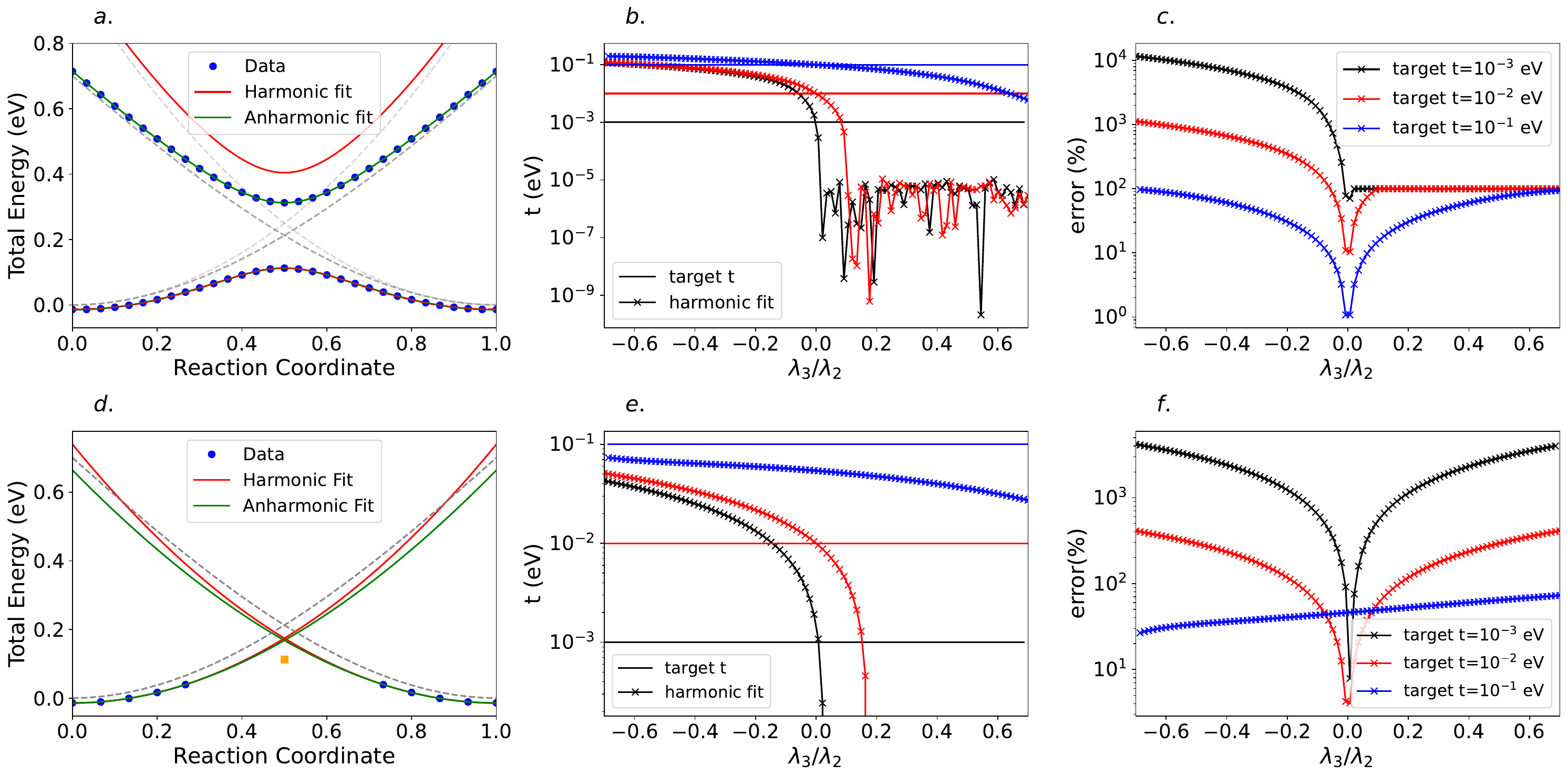}
    \phantomsubfloat{\label{fig:1a}}
    \phantomsubfloat{\label{fig:1b}}
    \phantomsubfloat{\label{fig:1c}}
    \phantomsubfloat{\label{fig:1d}}
    \phantomsubfloat{\label{fig:1e}}
    \phantomsubfloat{\label{fig:1f}}
\caption{
Influence on the extraction of coupling strength in two methods based on total energy: \protect\subref{fig:1a}-\protect\subref{fig:1c} ground state energy fitting method; \protect\subref{fig:1d}-\protect\subref{fig:1f} diabatic energy estimation method.
\textit{Ground state energy fitting method}: 
\protect\subref{fig:1a} Anharmonic data (blue dots) generated with parameters $\lambda_2 =1$~eV, $\lambda_3 = -0.3$~eV, and $t=0.1$~eV, is fitted using two-site model Hamiltonian  in harmonic approximation (in red) and with anharmonicity (in green). The grey transparent curve represents the diabatic curve obtained using equations \ref{eq:parabolic_Ea} and \ref{eq:parabolic_Eb} (harmonic approximation). The opaque grey curve is obtained using equation \ref{anharm_onsite_Ea} and \ref{anharm_onsite_Eb}. The lower eigenvalue obtained by harmonic fit is shown is a dashed line for visibility.  
\protect\subref{fig:1b} Extracted coupling strength from the model Hamiltonian in harmonic approximation (cross markers), as a function of the degree of anharmonicity ($\lambda_3/\lambda_2$) in data, shown for three different target values of the coupling constant. The target coupling strengths are shown by horizontal lines. 
\protect\subref{fig:1c} Log-scale plot of relative error, defined as  $|\frac{t_{\text{har.}} - t_{\text{tar.}}}{t_{\text{tar.}}}|$, in the extracted coupling strength in Harmonic approximation as a function of the degree of anharmonicity for three different coupling strengths. 
\textit{Diabatic energy estimation method:} 
\protect\subref{fig:1d} Harmonic (in red) and anharmonic (in green) fit for the diabatic curves over data generated in the same way as \protect\subref{fig:1a}. The gray curves represent the correct diabatic potential energy surface. The exact adiabatic energy at the transition state is shown by the orange square.
\protect\subref{fig:1e} and \protect\subref{fig:1f} are similar to \protect\subref{fig:1b} and \protect\subref{fig:1c}, but for the diabatic energy estimation method.
}
\label{fig:1}
\end{figure*}

The harmonic approximation in the two-site model (Eq.~\ref{eq:H_model}) is normally valid for energies close to the local minima. When the displacement of atoms is large (reaction coordinate deviates far from the equilibrium condition), anharmonicity can significantly distort the energy surface from the parabolic energy landscape shown in Fig.~\ref{fig:0}, as has already been reported in the context of Marcus theory in experimental and theoretical works \cite{small_theory_2003, yeganeh_effects_2006, tang_effects_1994}.
The presence of anharmonic effects hinders the accurate extraction of coupling strength. To demonstrate the influence of anharmonicity, we introduce a third-order term in the polaronic energy surface 
\begin{eqnarray}
    E_A &=& \lambda_A x^2 + \lambda_{A,3} x^3, \label{anharm_onsite_Ea} \\
    E_B &=& \lambda_B (1-x)^2 + \lambda_{B,3} (1-x)^3 + \Delta G \label{anharm_onsite_Eb},
\end{eqnarray}
For simplicity, we assume that the two polaron states are identical, with the same energy minimum ($\Delta G = 0$ eV), and energy surfaces are characterized by the same harmonic ($\lambda_2$) and anharmonic ($\lambda_3$) terms. We generated energy data points, which are the eigenvalues of the Hamiltonian described by Eq.~\ref{eq:H_model} in the presence of different anharmonicity $\lambda_3$ and coupling strength $t$ values, in order to test two approaches for coupling strength extraction based on the total energy approach. For simplicity, the harmonic term ($\lambda_2$) is fixed at $1$~eV in data generation. 

\subsection{Ground state energy fitting}
Since an $ab~initio$ calculation typically converges to the ground state, one approach to extract the coupling strength is to fit the ground state energies along the reaction coordinates using the lower eigenvalue of the model Hamiltonian  (Eq.~\ref{eq:H_model}). Figure \ref{fig:1a} shows the energy data (blue points) with anharmonicity $\lambda_3 = -0.3$~eV, and coupling $t=0.1$~eV. As expected, a model Hamiltonian  with anharmonicity can fit the ground state energy perfectly, predicting the exact value of parameters used in data generation. Therefore, the excited state is reproduced (green curve) by the model with the extracted parameters. If the model Hamiltonian  is under harmonic approximation, the ground state can also be fitted with high confidence ($99.96\%$) (dashed red curve). However, the extracted coupling strength is overestimated by $45.7\%$, i.e., $0.146$~eV $\pm 0.49\%$ instead of the exact value $0.1$~eV. This failure is also manifested in an overestimated excited state (full red curve).

To systematically assess the influence of anharmonicity on the coupling strength extraction, we selected three different coupling strength values; $t=10^{−1}$, $10^{−2}$ and $10^{−3}$~eV, and varied the anharmonicity in data with $\lambda_3 \in [−0.7,0.7]$~eV.
The target coupling constants are recovered if the model Hamiltonian  includes the anharmonic term, but it is not the case if the model Hamiltonian  is under harmonic approximation.
Fig.~\ref{fig:1b} presents the coupling strength extracted using the model Hamiltonian  under harmonic approximation by fitting $30$ data points of the ground state, which are equally spaced in reaction coordinates. Compared to correct coupling strengths (horizontal lines), the extraction with the fit is correct only when anharmonicity is absent in data, namely $\lambda_3 = 0$~eV.
Anharmonicity modifies the ground state energy profile, which  either shallows with $\lambda_3 < 0$ or steepens with $\lambda_3 > 0$ the energy barrier, resulting in an over- or under-estimation of the coupling strength. Especially for $\lambda_3 > 0$, the extracted coupling strength is way smaller compared to the target t. 
Fig.~\ref{fig:1c} quantifies the relative error on extracted $t$ values, indicating that weak coupling is difficult to reproduce. 
It is noticed that the relative error is not zero at $\lambda_3 = 0$ due to insufficient data points (30 points), which cannot resolve the signature of weak coupling at TS via this approach. 
Therefore, sufficient data points around TS are required to recover the target coupling strength.

Although anharmonicity drastically alters the total energy landscape, its effect on the ground-state total energy along the reaction coordinate can be disentangled from the influence of coupling strength. Therefore, the exact coupling strength can be extracted by including the anharmonic term in the model Hamiltonian  to fit the ground state energies as a function of reaction coordinates. \textit{ The inclusion of an anharmonic term in the model Hamiltonian  is necessary and not harmful even if anharmonicity is absent in the data.}

\subsection{Diabatic energy estimation}
To reduce the computational cost of obtaining multiple data points along the reaction path, an alternative method has been suggested to estimate the diabatic energy, $E^d_{TS}$ \cite{chen_tuning_2023,carey_hole_2021}. The coupling strength is extracted as the energy difference between this diabatic energy and the adiabatic ground state energy at the transition, expressed as $t = E^d_{TS} - E^-_{TS}$. The method follows three steps: 1) Fit the individual polaron state near its energy minimum, which only requires a few data points. 2) Determine the intersection by extrapolating the fitted energy surface to identify the diabatic state energy and the reaction coordinate at the transition. 3) Calculate the adiabatic ground state energy at the transition.

Figure~\ref{fig:1d} illustrates this method. The data points represent the lower eigenvalue of the Hamiltonian  (Eq.~\ref{eq:H_model}), which includes onsite energy with anharmonicity (Eq.~\ref{anharm_onsite_Ea} and \ref{anharm_onsite_Eb}). The parameters used to generate the data are the same as those in Fig.~\ref{fig:1a} ($\lambda_2 = 1$~eV, $\lambda_3 = -0.3$~eV, and $t = 0.1$~eV). The adiabatic energy at the transition, denoted as $E^-_{TS}$, is $0.112$~eV, represented by the orange square point.
Following the extraction procedure detailed in References \cite{chen_tuning_2023, carey_hole_2021}, we selected five data points for each polaron state near the energy minima. Specifically, we chose $x \in [0.0, 0.3]$ for the initial state and $x \in [0.7, 1.0]$ for the final state. These points were used to fit the energy profile while considering only the harmonic term, resulting in the red curves shown in Fig.~\ref{fig:1d}.
Although the data points close to the energy minima are fitted with a confidence level of $99.97\%$, the coupling strength is underestimated by $62\%$. The estimated diabatic energy, represented by the intersection of the red curves in Fig.~\ref{fig:1d}, is lower than the exact value indicated by the intersection of the gray curves. Furthermore, including an anharmonic term in the fit improves the situation only slightly, reducing the underestimation to $56\%$.

We conducted the same test as for the ground state fitting method by selecting three different coupling strengths: $t = 10^{-1} $, $ 10^{-2} $, and $10^{-3}$~eV and varying the anharmonicity parameter $\lambda_3$ in the range $[-0.7, 0.7]$~eV. The extraction procedure applied is consistent with the method described previously.
The extracted coupling strength, determined using the model Hamiltonian  under harmonic approximation, is presented in Fig.~\ref{fig:1e}. 
For strong coupling, indicated by $t = 0.1$~eV, the coupling strength is consistently underestimated (as shown by the blue curve), regardless of the sign of the anharmonic term. This underestimation occurs because the coupling strength lowers the ground state energy, even at the energy minima ( $x = 0$ or $1$ ). This reduction in energy results in a decreased estimated diabatic energy, leading to the observed underestimation of the coupling strength.
In the case of weak couplings, such as $t = 10^{-2}$ and $10^{-3}$~eV, the anharmonic term mainly perturbs the energy surface near the TS. Due to the absence of the third-order anharmonic term, the harmonic fit overestimates the diabatic energy for $\lambda_3 < 0$ , resulting in an overestimation of the coupling strength. Conversely, for $\lambda_3 > 0$, $t$ is underestimated. In case of large anharmonicity, the estimated diabatic energy is lower than the adiabatic energy, yielding a negative coupling constant. The only accurate extracted value occurs at $\lambda_3 = 0$~eV. 
Figure~\ref{fig:1f} quantifies the relative error for $t$, indicating that weak coupling strengths are challenging to reproduce since they are not on the same order of magnitude as the variations in total energy. It is observed that the lowest relative error occurs at $\lambda_3 = 0$ for weak couplings. However, this is not the case for strong coupling (indicated by $t = 0.1$~eV), as explained previously.

The polaron energy surface is affected by both coupling and anharmonicity. As a result, the method of fitting individual polaron energies using points near the energy minima can be affected by these two factors. \textit{This method fails to accurately estimate the diabatic energy, so it can not predict the correct coupling strength. Even if the anharmonic term is included, the strong coupling leads to incorrect predictions for the diabatic polaron energy surface and can mislead the determination of the coupling strength. Therefore, this method is not recommended.}

\raggedbottom

\section{Extraction from DFT+U}
\label{sec.DFT+U}
In this section, we present the extraction of coupling strength in the DFT+U framework. The aim is to highlight the influence of Hubbard $U$ on the typical physical picture (Fig.~\ref{fig:0}) in order to demonstrate the invalidity of coupling strength extraction by the model Hamiltonian  based on total energy. Instead, the method from the single particle electronic structure of neutral calculation is suggested as the reference method. We showcase three different hopping directions: $[00\bar{1}]$, $[100]$, and $[010]$ in V$_2$O$_5$ for illustration. 

\subsection{Destabilizing adiabatic state by Hubbard U: $[00\bar{1}]$}
\begin{figure*}
\includegraphics[width=2\columnwidth]{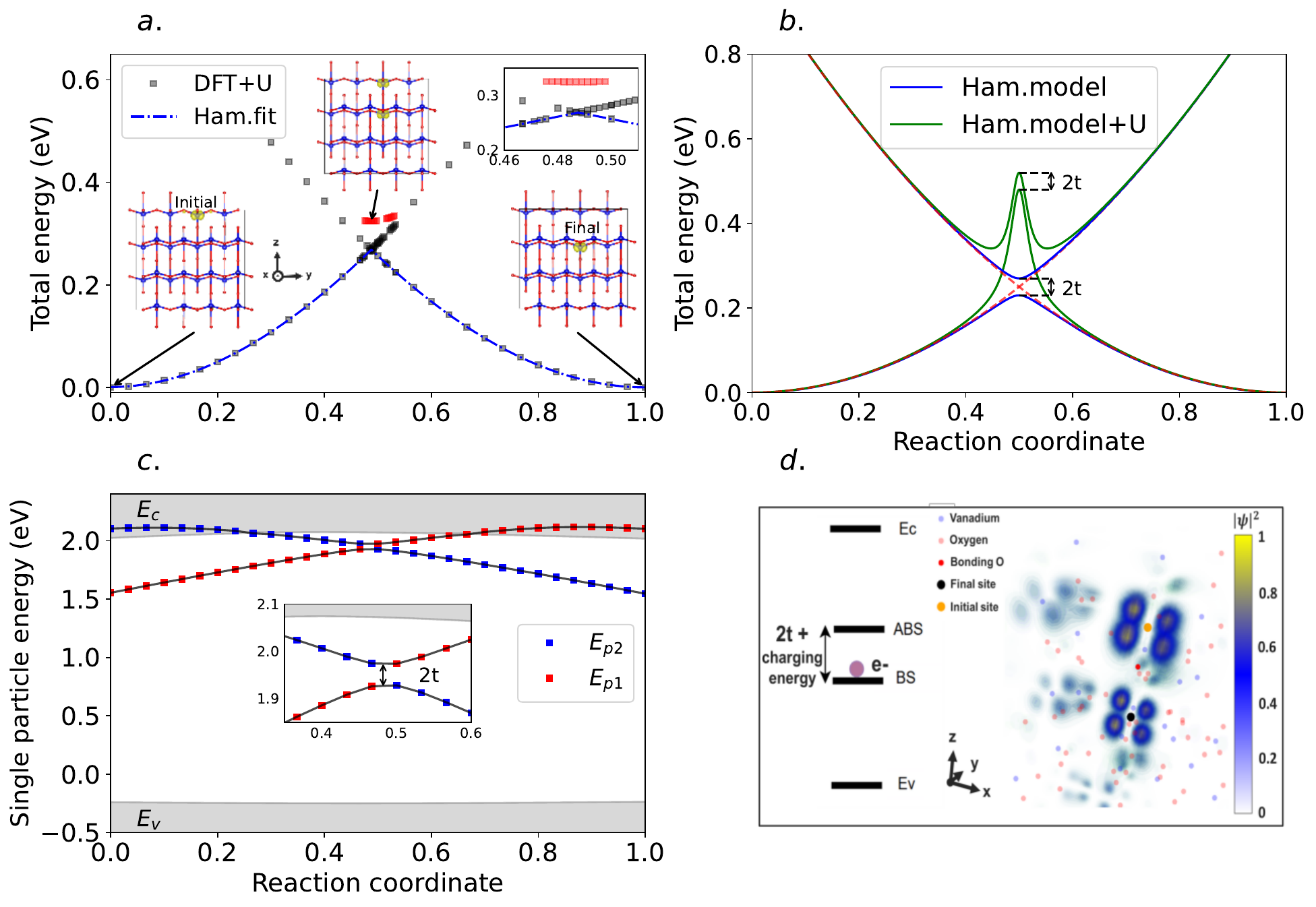}
\phantomsubfloat{\label{fig:2a}}
\phantomsubfloat{\label{fig:2b}}
\phantomsubfloat{\label{fig:2c}}
\phantomsubfloat{\label{fig:2d}}
\caption{
The polaron migration in the $[00\bar{1}]$ direction: \protect\subref{fig:2a} Total energy (gray squares) computed using DFT+U, showing diabatic character, is fitted by the model Hamiltonian  (in blue). A zoom around the TS is shown in the top right inset. The two bottom insets are the charge density of the initial and final states. An excited adiabatic branch around transition is also obtained from DFT+U (red squares), with polaron shared on two sites indicated by the top center inset.
\protect\subref{fig:2b} Influence of $U$ on total energy by comparing two-site model Hamiltonian  with (Eq.~\ref{eq:2site_model+Penalty}) and without $U$ (Eq.~\ref{eq:H_model}). In this example, we set $U=1$~eV, $\lambda_A=\lambda_B=1$ eV and $t=20$~meV. For simplicity, $\Delta G$ and anharmonicity are not considered. 
\protect\subref{fig:2c} Single-particle energy levels of neutral calculations (polaron level is unoccupied) along the reaction coordinates. Two polaron states are forming bonding and anti-bonding states at the transition (see inset).
\protect\subref{fig:2d} Charge density of polaron state $|\psi|^2$ at the transition obtained from single-particle wavefunction in charged calculation (the bonding state is occupied). Initial (orange) and final (black) vanadium atoms are linked by an oxygen atom (red).
}
\label{fig:2}
\end{figure*}

Fig.~\ref{fig:2a} presents the energy profile of the polaron hopping in the direction $[00\bar{1}]$. When a polaron forms, a single particle polaron state appears within the bandgap in the electronic structure. By examining the charge density of the single particle polaron state, we identify its charge localization on a vanadium atom at the initial or final state, as shown by the two bottom insets in Fig.~\ref{fig:2a}. DFT+U calculations of intermediate images, launched with the initial guess of charge density from that of the initial or final state, produce the energy landscape of each polaron state as a function of reaction coordinates. For $x<0.5$, starting with the charge density of the initial state, DFT+U yields the ground state. Meanwhile, it can converge to an excited state by starting with the charge density of the final state, when $x<0.5$. It is similar for $x>0.5$, where the final state charge density yields the ground state energies and the initial one yields the excited state. The two polaron states cross (see top right inset of Fig.~\ref{fig:2a}), showing a diabatic behavior, implying a small coupling strength. The coupling strength extracted using the model Hamiltonian  with anharmonicity is extremely small ($8.18 \times 10^{-10}$~eV), indicating no information of coupling strength is contained in the energy profile. Using default initialisation of charge density, we obtained another branch of an excited state (red points) with charge density equally shared on the two vanadium atoms, as shown by the top insert in Fig.~\ref{fig:2a}. The distribution of charge density shows an adiabatic behavior. However, we cannot obtain any adiabatic state with energy lower than the diabatic state at the transition. Such a result differs from the typical total energy profile presented in Fig.~\ref{fig:0}. Such a puzzle triggers us to ask a more fundamental question: \textit{How does the additional Hubbard $U$ influence the total energies of diabatic and adiabatic states?}

To answer this question, we introduce a Hubbard $U$ term to the model Hamiltonian , similar to DFT+U, i.e. the Dudarev implementation \cite{dudarev_electron-energy-loss_1998}: 
\begin{eqnarray}
\mathbf{H}_{U} = \mathbf{H} + U \sum_{I,\sigma} \mathbf{n^{I\sigma}(1-n^{I\sigma})}
= 
\begin{bmatrix}
E_A^U & t\\
t & E_B^U 
\end{bmatrix},
\label{eq:2site_model+Penalty} 
\end{eqnarray}
where the term $U \sum_{I,\sigma} \mathbf{n^{I\sigma}(1-n^{I\sigma})}$ represents the penalty applied on the total energy by summing over the sites $I$ and the spin channels $\sigma$. In the DFT+U calculation, $U$ is applied to all d-orbitals of Vanadium in V$_2$O$_5$. In contrast, $U$ is only applied to sites A and B in the model Hamiltonian $\mathbf{H_U}$. The onsite energy $E_A^U=E_A+Un_A(1-n_A)$ depends on the occupation of site A ($n_A$). A similar expression is used for the onsite energy of site B ($E_B^U$). Solving for the eigenstates of $\mathbf{H_U}$ is done in a self-consistent manner. This involves iterating through the process of determining the wavefunction $\psi$ by diagonalizing $\mathbf{H_U}$, computing the occupation $n=|\psi|^2$, and finally reconstructing $\mathbf{H_U}$. As an example, we set $U=1$~eV, $\lambda_A=\lambda_B=1$~eV and $t=20$~meV. For simplicity, $\Delta G$ and anharmonicity are not considered. 

Fig.~\ref{fig:2b} shows that this model $\mathbf{H_U}$ yields both adiabatic states ($E^-_{TS}$ and $E^+_{TS}$ states) above the diabatic intersection, which is in accordance with DFT+U observations (Fig.~\ref{fig:2a}). In fact, adiabatic states present partial occupations, raising large penalties to the total energy. Furthermore, the penalty varies with the occupation along the path. In contrast, diabatic states exhibit integer occupations, either $0$ or $1$, causing the penalty to vanish. This indicates that the penalty does not merely introduce a rigid shift and varies depending on the state considered. Consequently, the coupling strength can no longer be considered as the difference between the diabatic intersection and adiabatic point at the TS in the DFT+U framework, which has been employed previously \cite{chen_tuning_2023}. The correct coupling strength can only be extracted if one has access to both adiabatic ground and excited states in the total energy approach because both states should have similar $U$ correction due to a similar charge density. At the TS reaction coordinate, the BS of the $E^-$ ground state curve is occupied while the ABS is unoccupied. Conversely, the BS of the $E^+$ excited state curve is unoccupied while the ABS is occupied. The difference between those two total energy eigenvalues yield 2t (see green curve in Fig.~\ref{fig:2b}) .  In Fig.~\ref{fig:2a}, we were not able to capture the bump observed in the model Hamiltonian  $\mathbf{H_U}$ (cf. Fig.~\ref{fig:2b}). This can be attributed to the limitation of basis in the model Hamiltonian , which has only two sites (Eq.~\ref{eq:2site_model+Penalty}). In DFT+U, the basis spans real space, which has more degrees of freedom to minimize the total energy.

An alternative approach to extract coupling strength is from the electronic structure. Fig.~\ref{fig:2c} shows the evolution of single-particle energy of polaron states along the reaction coordinate. At $x=0$, the initial state $E_{p1}$ appears in the band gap, and the final state $E_{p2}$ has larger energy, so it is still buried inside the conduction band. The single-particle energy of the initial state rises with increasing $x$. Meanwhile, the energy of the final state decreases and appears in the band gap at $x\approx 0.3$. At the TS $x=0.5$, the two polaron states are at the resonant condition, forming single-particle bonding and anti-bonding states, shown in Fig.~\ref{fig:2d}. The energy difference between the two states is twice the coupling strength. The energy of the final state continues to decrease with larger $x$, and the initial state with an increase in energy enters the conduction band at $x\approx0.7$.

In principle, one needs only to evaluate the electronic structure at TS to extract the coupling strength. However, attention has to be paid to two points when using this method. 1) Clearly identifying the bonding and anti-bonding states is necessary. Generally, the bonding state stays in the gap. The anti-bonding state should decrease its energy, reach a minimum at the TS, and appear in the bandgap. 
2) In $ab~initio$ calculation for the polaron state, the system typically has an additional charge. Therefore, the bonding state is occupied, but the anti-bonding state is unoccupied at the TS. The additional charge in the bonding state introduces extra charging energy for the anti-bonding state, resulting in a more significant energy splitting, which is not purely due to the coupling strength (cf Fig.~\ref{fig:2d}). Furthermore, the additional charge under periodic boundary conditions requires extra correction, such as the FWP scheme \cite{falletta_finite-size_2020}. Therefore, extracting the coupling strength in the charge-neutral condition is more physically plausible, i.e., both bonding and anti-bonding states are unoccupied for electron-polaron, as presented in Fig.~\ref{fig:2c}. For polaron hopping in $[00\bar{1}]$, the coupling strength is about $20$~meV in the neutral condition, while it is overestimated at $141$ meV with additional charge due to the contribution of charging energy to the splitting. 


\subsection{In case of large coupling strength: $[100]$}

\begin{figure*}
\includegraphics[width=2\columnwidth]{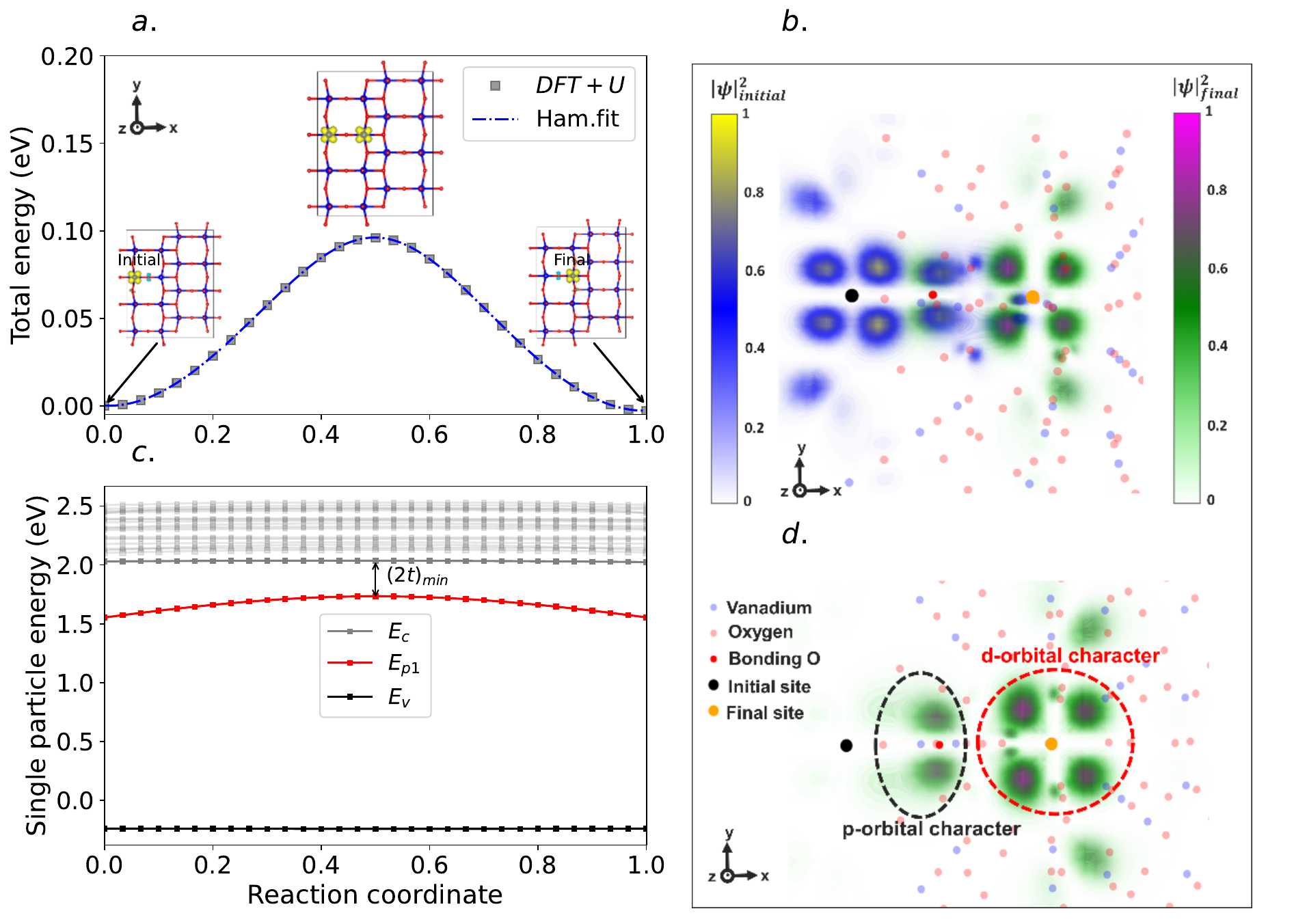}
\phantomsubfloat{\label{fig:3a}}
\phantomsubfloat{\label{fig:3b}}
\phantomsubfloat{\label{fig:3c}}
\phantomsubfloat{\label{fig:3d}}
\caption{
The polaron migration in the $[100]$ direction: 
\protect\subref{fig:3a} Total energy (gray dots) computed using DFT+U showing adiabatic character is fitted by the model Hamiltonian . The charge density of the initial, transition and final states are shown by insets.
\protect\subref{fig:3b} Charge density of initial and final polaron states overlaid at the transition state atomic positions. Charge density $|\psi|^2$ is obtained from a single-particle polaron state in charged calculations. Initial (black) and final (orange) vanadium atoms are linked by an oxygen atom (red).
\protect\subref{fig:3c} Single-particle energy levels of neutral calculations (polaron level is unnocupied) along the reaction coordinates. Only single polaron state (in red), which corresponds to the bonding state at the transition, is identified in the gap. The anti-bonding state can not be identified. $E_v$ represents the maximum of the valence band and $E_c$ the minimum of the conduction band.
\protect\subref{fig:3d} The charge density of the final polaron state shows a hybridization of the d-orbital character from the final vanadium atom and the p-orbital of the oxygen atom in between. 
}
\label{fig:3}
\end{figure*}

Fig.~\ref{fig:3a} reports the polaron hopping in the direction $[100]$ of V$_2$O$_5$. Unlike diabatic transition in $[00\bar{1}]$, DFT+U yields an adiabatic transition with a smoother barrier. By testing various initial charge densities and magnetic momentum, the calculations did not converge to any excited state. Having noticed the influence of U, which destabilizes adiabatic states, we wondered \textbf{why the adiabatic solution is more stable for this specific jump.}

By analyzing the density of the polaron single-particle state $|\psi|^2$, we noticed strong hybridization between the d-orbital of vanadium and p-orbital of oxygen atoms, as shown by Fig.~\ref{fig:3d} for the final state. The hybridization reduces the penalty on total energy since $U$ only applies on d-orbitals of Vanadium. Furthermore, the initial and final states show significant overlap on the p-orbital of the same oxygen atom, as shown in Fig.~\ref{fig:3b}, implying a considerable coupling strength to stabilize the adiabatic state. By employing the model Hamiltonian  with anharmonicity, we obtain a coupling strength of $124$~meV.

In Fig.~\ref{fig:3c}, we report the single-particle energy levels along the polaron migration path in the neutral conditions. We notice the absence of the anti-bonding state in the gap. Because of the considerable coupling strength, the anti-bonding state is pushed inside the conduction band and mixed with other states. Such a finding is consistent with the significant overlap between the initial and final single-particle states as shown in Fig.~\ref{fig:3b}. Due to the absence of the anti-bonding state in the band gap, the value extracted between the bonding state and the minimum of the conduction band at the TS will correspond to a minimal coupling strength of $t=150$~meV, larger than $124$~meV extracted from the model Hamiltonian . Therefore, the value obtained from the band structure should be regarded as the best one we can extract.

\subsection{Presence of a third state: $[010]$}
\begin{figure}
\includegraphics[width=0.45\textwidth]{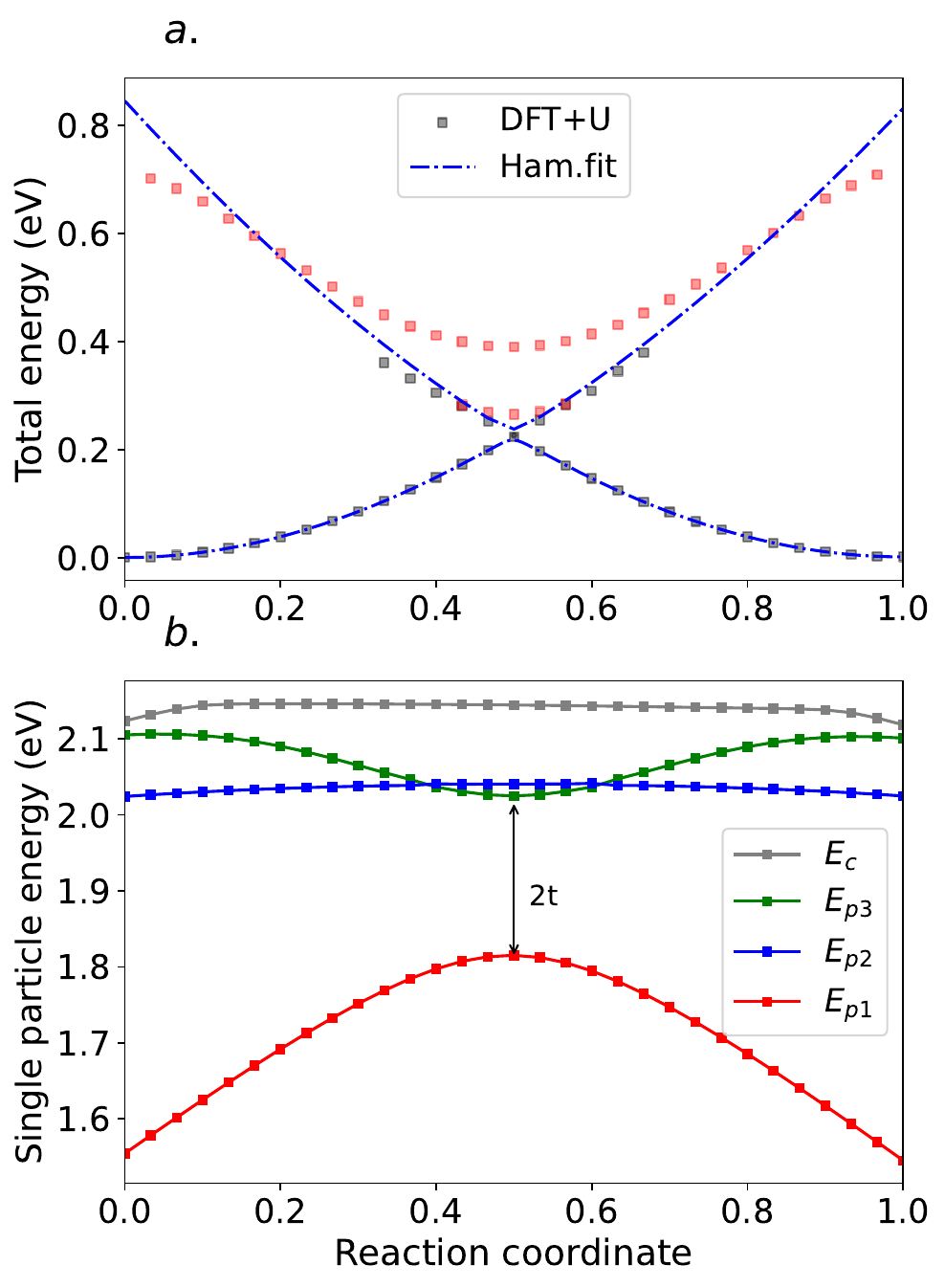}
\phantomsubfloat{\label{fig:4a}}
\phantomsubfloat{\label{fig:4b}}
\caption{The polaron migration in the $[010]$ direction: \protect\subref{fig:4a} Total energy (gray dots) computed using the DFT+U, showing diabatic character. Two branches of excited adiabatic states (red dots) demonstrate the presence of a third state in the transition. A two-site model Hamiltonian  is unable to reproduce the DFT+U results. 
\protect\subref{fig:4b} Single-particle energy levels of neutral calculations (polaron level is unoccupied) along the reaction coordinates, demonstrate the presence of a third state close to the transition, in accordance with total energy surfaces. $E_c$ represents the next higher band in energy following the third polaron state.}
\label{fig:4}
\end{figure}

Fig.~\ref{fig:4a} reports the polaron hopping in the direction $[010]$ of V$_2$O$_5$. Similar to $[00\bar{1}]$ direction, the diabatic state is the most stable at TS. However, we were able to stabilize two distinct adiabatic excited states above this diabatic intersection (two red branches). The evolution of these bands demonstrates the involvement of the third state in the transition. 
Those three states are also displayed in the band structure approach (Fig.~\ref{fig:4b}). As the reaction coordinate increases from the initial state, an excited state, denoted in green, decreases in energy and intersects with another excited state in blue, which is flat in energy. The intersection at $x\approx0.4$ (similar for $x\approx 0.6$) between the two excited states does not show significant anti-crossing, with a coupling strength of about $5$~meV. Meanwhile, the bonding state increases in energy, reaching a maximum at $x=0.5$, which is the TS. From the energy splitting between the bonding and anti-bonding states, we extract the coupling strength of $t=105$~meV. 

The presence of the third state is due to immense coupling strength in the direction $[100]$. Although the reaction coordinate is for polaron hopping along $[010]$ direction, the energy of the initial state is not only approaching the final state in $[010]$ direction but also the excited state localized on the neighboring atom in $[100]$ direction for $x<0.5$. Similarly, when $x>0.5$, the neighboring atom in the $[100]$ direction relative to the atom where the final state is localized intervenes. This case highlights that polaron hopping involves not merely two sites; other sites with considerable coupling strength can also influence the hopping process. Besides checking only the band edge (lowest conduction band or highest valence band), it is essential to verify deeper bands when the band structure method is employed, as they can provide more information about the coupling strength between involved states. 

\section{Extraction from HSE06 Hybrid Functional}
\label{sec.HSE}
\begin{figure*}
\includegraphics[width=2.1\columnwidth]{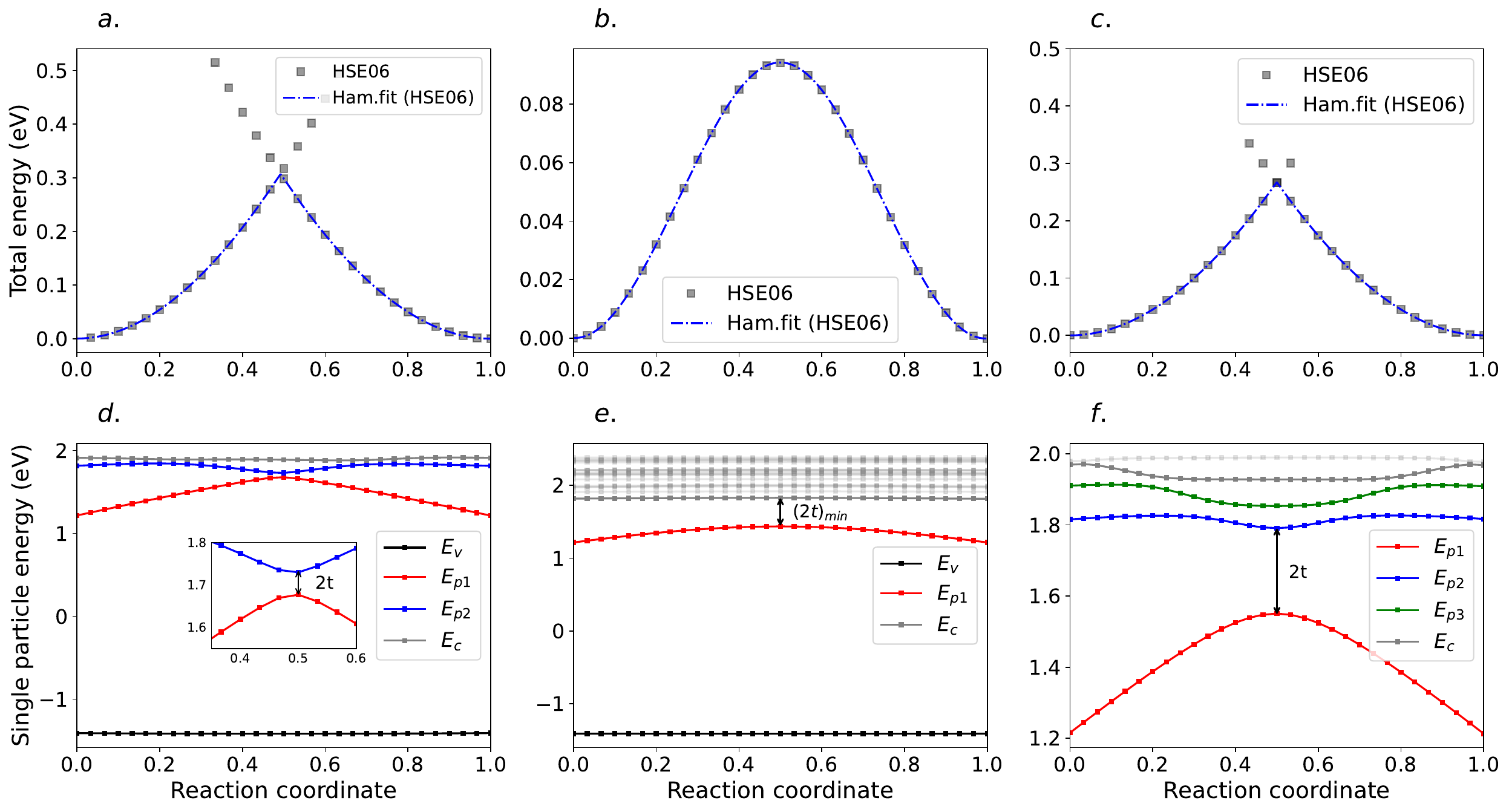}
\phantomsubfloat{\label{fig:5a}}
\phantomsubfloat{\label{fig:5b}}
\phantomsubfloat{\label{fig:5c}}
\phantomsubfloat{\label{fig:5d}}
\phantomsubfloat{\label{fig:5e}}
\phantomsubfloat{\label{fig:5f}}
\caption{Polaron migration along $[00\bar{1}]$ \protect\subref{fig:5a},\protect\subref{fig:5d}, $[100]$ \protect\subref{fig:5b},\protect\subref{fig:5e} and $[010]$ \protect\subref{fig:5c},\protect\subref{fig:5f} directions computed using HSE06 functional. \protect\subref{fig:5a}-\protect\subref{fig:5c} Total energy (gray dots) from HSE06 calculations are fitted using the model Hamiltonian  (blue). \protect\subref{fig:5d},\protect\subref{fig:5f} Single-particle energy levels of neutral HSE06 calculations (polaron level is unoccupied).
The HSE06 results qualitatively agree with DFT+U.  
}
\label{fig:5}
\end{figure*}

We repeated the study of DFT+U reported in section~\ref{sec.DFT+U} by using HSE06 hybrid functional, although it requires more computational resources. Figure~\ref{fig:5} presents the results, which are qualitatively consistent with DFT+U, implying our findings are not limited in the framework of DFT+U. Fig.~\ref{fig:5a} and Fig.~\ref{fig:5c} indicate the stabilization of the diabatic state in $[00\bar{1}]$ and $[010]$. The coupling strength can only be extracted from electronic structure, as bonding and anti-bonding states are identified in the band-gap, shown in Fig.~\ref{fig:5d} and Fig.~\ref{fig:5f}. The adiabatic state is stabilized in $[100]$ direction (Fig.~\ref{fig:5b}), because of a strong coupling strength. The anti-bonding state is pushed inside the conduction band and can not be identified from the electronic structure (Fig.~\ref{fig:5d}). Therefore, the coupling strength from the energy splitting gives only a lower bound. 

\section{Discussion}\label{sec.discuss}
Table~\ref{tab:t_results} summarizes the extracted coupling strengths obtained from fitting the ground state total energy using the model Hamiltonian, as well as from the energy splitting between bonding and anti-bonding states in electronic structure calculations (charged and neutral case), for both DFT+U (section~\ref{sec.DFT+U}) and HSE06 (section~\ref{sec.HSE}). In the table, the coupling strengths are in color, indicating trustable (blue), correct under certain limitations (orange), and incorrect (red) results. In the following, we discuss each method individually.

\begin{table*}
    \caption{Comparison of coupling strength of polaron hopping extracted (in eV) using various methods in three directions in V$_2$O$_5$ using DFT+U or HSE06 hybrid functional. The total energy method is fitting ground state energy along the reaction path discussed in section~\ref{sec.TE}. The electronic structure method uses the energy splitting between bonding and anti-bonding state in electronic band-structure, mentioned in section~\ref{sec.ES}.The extracted coupling strength is written in the color code, indicating trustable (blue), correct under certain limitations (orange), and incorrect (red) results. $q=-1$ corresponds to the charged calculation and $q=0$ corresponds to the neutral one. 
}
    \label{tab:t_results}   
    \begin{ruledtabular}t (eV)
        \begin{tabular}{lcccccc}
            \multicolumn{1}{c}{\multirow{3}{*}{Direction}} & \multicolumn{3}{c}{DFT+U} & \multicolumn{3}{c}{HSE06} \\
            \cline{2-4} \cline{5-7}
            & \multirow{2}{*}{Total Energy} & \multicolumn{2}{c}{Electronic Structure} & \multirow{2}{*}{Total Energy} & \multicolumn{2}{c}{Electronic Structure} \\
            \cline{3-4} \cline{6-7}
            & & $q=-1$ & $q=0$ & & $q=-1$ & $q=0$ \\
            \hline
            $[00\bar{1}]$ & \textcolor{red}{$8.18\times10^{-10}$} & \textcolor{red}{0.141} & \textcolor{blue}{0.020} & \textcolor{red}{$1.0 \times10^{-3}$} & \textcolor{red}{0.296} & \textcolor{blue}{0.026} \\
            $[100]$       & \textcolor{orange}{$0.124$} & \textcolor{red}{0.354} & \textcolor{orange}{0.150} & \textcolor{orange}{$0.361$} & \textcolor{red}{0.542} & \textcolor{orange}{0.196} \\
            $[010]$       & \textcolor{red}{$0.010$} & \textcolor{red}{0.255} & \textcolor{blue}{0.105} & \textcolor{red}{$5.72\times 10^{-4}$} & \textcolor{red}{0.405} & \textcolor{blue}{0.120} \\
        \end{tabular}
    \end{ruledtabular}
\end{table*}

\subsection{Total energy method}
\label{sec.TE}
The extraction of coupling strength from the total energy of the system along the reaction path, either by fitting the ground state energy or by estimating the diabatic energy from the polaron energy surface extrapolation, is based on a simple two-site model Hamiltonian , which is typically under the harmonic approximation. In the presence of anharmonicity, which is physical and significant at large atomic displacement with strong electron-phonon coupling, such a model would lead to incorrect coupling strength, as demonstrated in section~\ref{sec.anharmonicity}.
Including an anharmonic term in the model Hamiltonian  improves the accuracy of coupling strength extraction with the ground state fitting method. Still, it does not help with the diabatic energy estimation method.

More severely, Coulomb interaction is missing in the model Hamiltonian \textbf{$H$} , which approximates the coupling strength as being the energy difference of the diabatic state to the ground adiabatic state, i.e., $t= E^{d}_{TS}-E^-_{TS}$ (see Fig.~\ref{fig:0}). This is incorrect because the charge density differs in the adiabatic and diabatic states. The Coulomb interaction would affect the energy difference between the diabatic and adiabatic states so that the energy splitting is not purely a result of the coupling strength in $ab~initio$ calculations.
With Coulomb interaction, calculations tend to stabilize a more localized diabatic state. The diabatic state becomes a ground state at weak coupling strength. In such a case, the total energy of the ground state does not contain any information on coupling strength, as demonstrated by hopping along $[00\bar{1}]$ and $[010]$ directions. Extracting coupling strength requires access to the ground and excited adiabatic states at the TS. Because these two states have similar charge densities, the energy splitting between them can be considered twice the coupling strength (see Fig.~\ref{fig:2b}). However, both states are now excited and are not easily accessible by DFT. If the adiabatic state is stabilized, the coupling strength can be extracted by the model Hamiltonian in the limit of the absence of Coulomb interaction, as demonstrated by the case of $[100]$ where DFT+U and HSE show a large discrepancy (reported in Table~\ref{tab:t_results}) using the model Hamiltonian due to the different description of coulomb interaction by the two functionals.

\subsection{Electronic structure method}
\label{sec.ES}
Polaron hopping is the transition between two polaron states. The initial state rises in energy, and the final state decreases. At the transition point, these two states are resonant. The coupling strength hybridizes the two states, forming the bonding and anti-bonding adiabatic states, with the energy splitting twice the coupling strength. Therefore, the coupling of the polaron states is electronic, and its extraction requires only the atomic structure at TS. Moreover, a polaron state localized in space is well described by a single-particle state falling in the bandgap of the electronic structure. At TS, the bonding and anti-bonding single-particle polaron states result from coupling strength, and their energy difference is twice the coupling strength only if the two states are both occupied for hole-polaron or unoccupied for electron-polaron. Such occupation is the neutral condition for a crystal and is chosen to avoid additional energy splitting due to Coulomb interaction, which would result in an overestimation of the coupling strength. The additional energy, known as the charging energy \cite{shokrani_size-dependent_2025}, is the extra energy required to add an electron to the anti-bonding state when the bonding state is occupied. The overestimation of coupling strength in charged condition ($q=-1$) is shown in Table~\ref{tab:t_results} for both DFT+U and HSE06 results for all three hoppings. The coupling strength extracted in neutral condition agrees better between DFT+U and HSE06.

\section{Conclusion} 
\label{sec:conclusion}
In this work, we highlight the importance of including anharmonicity in the model Hamiltonian by estimating the error on the extracted coupling strength via the total energy methods. Fitting the ground state energy along the reaction coordinate based on a model Hamiltonian  including an anharmonic term, is necessary and not harmful even if anharmonicity is absent in data. However, fitting individual polaron energy surfaces, even with anharmonicity, then estimating coupling strength from the energy difference between diabatic and adiabatic states is questionable.



Furthermore, we revisit different approaches to extract the coupling strength by studying three different jumps in V$_2$O$_5$ using DFT+U and HSE06. We mainly demonstrate the limitations of employing the total energy method to extract the coupling strength. We show that $ab~initio$ method tends to stabilize diabatic solution for weak coupling strength due to coulomb repulsion, which is not considered in the two site model Hamiltonian . This diabatic barrier does not contain any information on coupling strength and can not be used for extraction. For the same reason, extracting the coupling strength as the energy difference between diabatic and adiabatic states is erroneous. A sanity check consists of extracting the adiabatic bonding and anti-bonding state at the TS. In those steps, we present an alternative approach by extracting $t$ from both unoccupied bonding and anti-bonding states for polaron electron in electronic structure, to eliminate the influence of the charging energy in energy splitting at TS. For completeness, We also highlight the limitation of band structure extraction in the strong coupling case. 




\begin{acknowledgments}
This project is financed by the French government’s France 2030 investment program and the french national agency of research (ANR-CIFRE n°2023/0835). Part of the calculations were performed on computational resources provided by GENCI–IDRIS (Grant 2025-A0170912036).

\end{acknowledgments}

\appendix

\section{Hubbard $U$  parameter for polaronic defects}
\label{detU}

The Hubbard $U$ parameter used throughout this work was fixed to correct for Self-Interaction Error (SIE) by satisfying piecewise linearity of the total energy as function of  occupation, a known property of exact functionals \cite{kronik_piecewise_2020}. This was achieved by using the method of Falletta et al. \cite{falletta_hubbard_2022}, specially developed for polaronic defects. In the latter approach, using Janak's theorem \cite{janak_proof_1978}, Hubbard $U$ should be fixed so that the eigenvalue of the polaron $\epsilon_p$ remains constant with the change of occupation (charge $q$), namely: 
\begin{equation}
    \dfrac{d}{dq} \epsilon_p^U(q) = 0 \Longrightarrow \epsilon_p^U(0)= \epsilon_p^U(-1)
    \label{falleta}
\end{equation}

This is achieved by computing the energy level of a polaronic state for multiple values of $U$ and for two different charge states, specifically $q=0$ and $q=−1$. Subsequently, the $U$ value is determined when the two eigenstates are equal for different occupations as depicted by Eq.~\ref{falleta}. The $U$ value extracted for V$_2$O$_5$ is $U \approx 3.5 $ eV. Note that this values remains very close to the one determined by Wang et al. \cite{wang_oxidation_2006} ($U=3.25$ eV) by matching the experimental enthalpies of redox reaction of Vanadium oxides. 

\begin{figure}
\includegraphics[width=1\columnwidth]{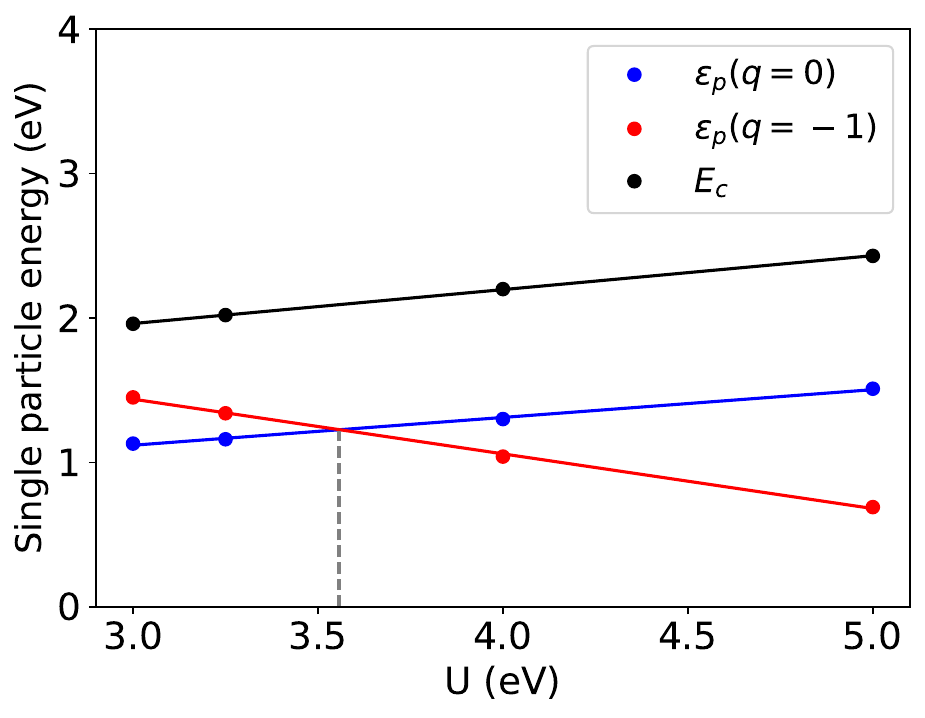}
\caption{ Energy of single-particle polaron state $\epsilon_p^U(q)$ as function of $U$ value for the neutral (unoccupied in blue)  and charged (occupied in red) cases. The intersection between the two determine the $U$ value, which satisfies the piecewise linearity of the total energy for the system. The $U$ value is calculated to be around $3.5$~eV for polaron states in V$_2$O$_5$.}
\label{fig:S1}
\end{figure}

Since periodic boundary conditions are applied, we use finite size corrections to correct for the replications of the charge and the distortion using FWP (Falleta, Wiktor, Pasquarello) scheme \cite{falletta_finite-size_2020}. The correction added to the total energy of a defect state of charge $q$ is defined as : 
\begin{equation}
\begin{aligned}
E_{\text{cor}}(q, R_{q'}) &= E_{\text{m}}(q', \varepsilon_0) 
- E_{\text{m}}(q' + q'_{\text{pol}}, \varepsilon_\infty) \\
&\quad + E_{\text{m}}(q + q'_{\text{pol}}, \varepsilon_\infty)
\end{aligned}
\end{equation}
The structure containing this charged defect $q$ is relaxed in the presence of the charge $q'$ and is denoted $R_{q'}$. These corrections employ the FNV (Freysoldt, Neugebauer, and Van
de Walle) scheme \cite{freysoldt_electrostatic_2011} via three terms. The first two terms account for ionic polarization $q'_\text{pol}$ by using the static dielectric constant $\epsilon_0$ and the electronic dielectric constant $\epsilon_{\infty}$. The last term accounts for the electronic response to the localised charge  $q + q'_\text{pol}$.  The correction applied to the single particle energy is \cite{falletta_finite-size_2020}:
\begin{equation}
\varepsilon_{\text{cor}}(q, \mathbf{R}_{q'}) = -2 \frac{E_{\text{m}}(q + q'_{\text{pol}}, \varepsilon_\infty)}{q + q'_{\text{pol}}}.
\end{equation}

The dielectric constants are determined as directional averaged value from the diagonal of the ionic and electronic dielectric tensors computed using VASP. The high-frequency dielectric electronic constant is $\epsilon_{\infty} = \mathrm{Tr}(\Sigma_{elec})/3$; and the static dielectric constant is $\epsilon_{0} = \mathrm{Tr}(\Sigma_{elec}+\Sigma_{ion})/3$.
We determined the dielectric constant at four different $U$ values ($U= 3$, $3.25$, $4$, $5$~eV), the results are very similar, with small discrepancy: $\epsilon_{\infty}=5.87 \pm 0.01$ and $\epsilon_{0}=20.0 \pm 0.27$. The extracted dielectric constants remain reasonably close to experimental work ($\epsilon^{EXP}_{\infty}=4.22$ \cite{n_kenny_optical_1966} and $\epsilon^{EXP}_{0}=13.8$ \cite{lamsal_optical_2013}). Additionally, the corrections are not very sensitive to the choice of the dielectric constant when it is large, as mentioned by Windsor et al. \cite{windsor_treating_2024}. 



\nocite{*}

\bibliography{coupling_extraction}

\end{document}